\documentclass[ajl]{emulateapj}

\usepackage{graphicx,epsfig,natbib,color}
\usepackage{lscape}
\usepackage{graphicx}
\usepackage{epsfig}
\usepackage{natbib}
\usepackage[colorlinks, citecolor=blue]{hyperref}
\definecolor{DarkGreen}{rgb}{0.0,0.45,0.0}  % define a custom color

\slugcomment{Received 2018 January 7; accepted 2018 February 13}

\begin{document}
\title{Unambiguous Evidence of Filament Splitting-Induced Partial Eruptions}

\author{X. Cheng$^{1,2}$, B. Kliem$^{3}$, \& M. D. Ding$^{1,2}$}

\affil{$^1$School of Astronomy and Space Science, Nanjing University, Nanjing 210093, China}\email{xincheng@nju.edu.cn}
\affil{$^2$Key Laboratory for Modern Astronomy and Astrophysics (Nanjing University), Ministry of Education, Nanjing 210093, China}
\affil{$^3$Institute of Physics and Astronomy, University of Potsdam, D-14476 Potsdam, Germany}

\begin{abstract}
Coronal mass ejections are often considered to result from the full eruption of a magnetic flux rope (MFR). However, it is recognized that, in some events, the MFR may release only part of its flux, with the details of the implied splitting not completely established due to limitations in observations. Here, we investigate two partial eruption events including a confined and a successful one. Both partial eruptions are a consequence of the vertical splitting of a filament-hosting MFR involving internal reconnection. A loss of equilibrium in the rising part of the magnetic flux is suggested by the impulsive onset of both events and by the delayed onset of reconnection in the confined event. The remaining part of the flux might be line-tied to the photosphere in a bald patch separatrix surface, and we confirm the existence of extended bald-patch sections for the successful eruption. The internal reconnection is signified by brightenings in the body of one filament and between the rising and remaining parts of both filaments. It evolves quickly into the standard current sheet reconnection in the wake of the eruption. As a result, regardless of being confined or successful, both eruptions produce hard X-ray sources and flare loops below the erupting but above the surviving flux, as well as a pair of flare ribbons enclosing the latter.
\end{abstract}

\keywords{Sun: corona --- Sun: coronal mass ejections (CMEs) --- Sun: magnetic fields --- Sun: flares}
\clearpage

\section{Introduction}

Coronal mass ejections (CMEs) and solar flares are the most energetic explosions in the solar system. They release large quantities of magnetized plasma, electromagnetic radiation ranging from decameter radio waves to $\gamma$ rays, and energetic particles into the interplanetary space \citep{forbes06,benz08,chen11_review,schmieder15,cheng17_review}. When these products arrive at the Earth, they interact with the magnetosphere and ionosphere, causing magnetic or ionospheric storms that can impact the safety of human high-tech activities \citep{webb00}. 

The pre-eruptive configuration of CMEs/flares is often believed to be a magnetic flux rope (MFR), which may contain a filament/prominence \citep{kuperus74,vantend78,vanballegooijen89,mackay10,schmieder13,cheng17_review}. In this system, filament material is preferentially collected in the magnetic dips of the MFR, which provide an upward Lorenz force that naturally balances gravity \citep{low95_apj,gibson04,gibson06_jgr}. Such a picture actually has been confirmed by many observations such as the descending motion of filament material along a helical trajectory \citep{liting13,cheng14_tracking,joshi14,zhangjun15}, spinning motions in filament channels \citep{gibson04,wangym10,lixing12}, solar tornadoes \citep{zhangjun11,wedemeyer12,suyang12_tornadoes,suyang14_tornadoes}, and untwisting motions in erupted filaments \citep{yanxl14,xuezhike16}. In addition, through non-linear force-free field (NLFFF) modelling, it was found in many cases that the dips of twisted field lines are cospatial with the filament locations \citep[e.g.,][]{aulanier00_filament,guo10_filament,cheng13_double,suyingna11,jiang14_nlfff,yanxl15,biyi15}.

Theoretically, it is thought that the loss of equilibrium of an MFR leads to eruption \citep{forbes06,chen11_review,schmieder15}. The evolution toward the loss of equilibrium most likely involves magnetic reconnection that forms an MFR or facilitates its formation \citep{chifor06,chifor07,tripathi09}. This is referred to as tether-cutting reconnection if occurring in and below the MFR \citep{moore01} and as breakout reconnection if occurring at a high-lying X-line \citep{antiochos99}. The tether-cutting reconnection primarily increases the upward magnetic pressure below the MFR, while the breakout reconnection primarily reduces the downward magnetic tension of overlying flux. The loss of equilibrium can be described as ideal magnetohydrodynamic (MHD) instability of an MFR, including the torus instability \citep{kliem06,olmedo10} and the helical kink instability, in the following simply referred to as the kink instability \citep{hood81,torok04,fan04,srivastava10}. Equivalently, it represents the catastrophe of an MFR in parameter space \citep{forbes91,linjun00,linjun02}. The catastrophes investigated in the context of CMEs and flares have been demonstrated to be equivalent to the torus instability \citep{demoulin10,kliem14_torus}. Their occurrence requires that the restoring force of the field due to sources external to the MFR (background field) declines sufficiently rapidly with height. The kink instability occurs if the MFR is sufficiently twisted; it transforms the twist into writhe, i.e., a helical shape of the MFR axis. Observationally, the features such as the inverse-$\gamma$ shape configuration \citep{ji03,williams05,guo10_index,yanxl14}, rotation of the filament axis about the direction of ascent \citep{rust05,gilbert07,green07}, and reconnection at the crossing point of the filament legs \citep{liurui09,kliem10,tripathi13} have been considered to be evidence of the kink instability.

In describing CMEs and flares, it is often assumed that an MFR erupts as a whole. However, the source region quite often undergoes a partial eruption, which releases only part of the flux. \citet{gilbert00} examined 54 H$\alpha$ prominences and filaments and found that a majority of the 18 eruptive prominences in their sample exhibited a separation into escaping and remaining material in the range of projected heights of $0.1\mbox{--}0.4~R_\odot$, i.e., even the remaining material showed a rise up to this height range. Typically, the bulk of the prominence fell back to the solar surface \citep[for a similar case, see][]{zhangqm15}. Sometimes, the downflowing filament materials also brightened, even emitting in the X-ray passband, indicating heating at their sources \citep{tripathi-solanki06,tripathi07}. The disconnection was inferred to be due to reconnection taking place in the middle of an MFR. \citet{gilbert01} further conjectured that an MFR should be totally expelled if reconnection occurs below it and partially expelled if the reconnection happens internally. 

More recent observations have revealed that many eruptions leave a filament in the source region essentially unchanged, i.e., the flux holding the filament does not appear to rise at all. Often a hot structure, a soft X-ray sigmoid or an EUV hot channel, is seen to erupt above a remaining filament \citep{pevtsov02, cheng14_formation,dudik14}. In other cases, vertically split filaments are observed whose upper part erupts \citep{suyingna11, liurui12_filament}. To accommodate this behaviour, \citet{gibson06_apjl} extended the scenario put forward in \citet{gilbert00, gilbert01} by suggesting that the erupting structure is an MFR with a bald patch (BP). A BP is a section of the photospheric polarity inversion line (PIL) where the separatrix surface at the outer edge of the MFR touches the photosphere tangentially. The separatrix surface is then referred to as a bald-patch separatrix (BPS). The photospheric line tying in the BP prevents the lower part of the MFR from erupting and thus provides a natural explanation for the vertical splitting of MFRs whose upper part is unstable. The splitting involves reconnection within the MFR. This mechanism for partial eruptions was demonstrated through numerical simulation by \citet{gibson06_apjl, gibson08_jgr}. 

A kink-unstable MFR may also split into an escaping and a remaining part if situated above an arcade structure, i.e. in the absence of BPs, as demonstrated by \citet{birn06}. However, the parameters of the initial configurations in these simulations were chosen such that the instability commenced only for a high twist of at least 4.5 turns, which is much higher than the twists typically inferred in coronal source regions prior to an eruption. Whether this configuration allows for a splitting at lower twist is not yet clear.

Subsequently, \citet{tripathi09_partial} observed many of the features predicted by \citet{gibson06_apjl}, such as surviving filament material, flare ribbons on either side of the surviving filament (which indicate internal reconnection), and the transformation of the associated sigmoid to a cusp-over-sigmoid structure \citep[also see][]{tripathi13}. However, although all of these features are expected results of partial eruptions, the core process suggested in \citet{gilbert00}---that an originally single filament-hosting MFR is vertically split along its axis into two parts by internal reconnection and only the upper flux is ejected---has never been clearly observed for a filament. Only two of the six cases studied in \citet{tripathi09} reveal some aspects of that scenario unambiguously (weak indication of a remaining filament but missing signs of reconnection along most of its length on 2003 Aug 25; and flare ribbons enclosing a remaining filament on 2000 Sep 12). The other cases do not allow distinguishing between a remaining and a reformed filament, due to the large data gaps and narrow temperature sensitivity of the employed H$\alpha$ observations. 

\begin{figure*}
\centering {\includegraphics[width=16.9cm]{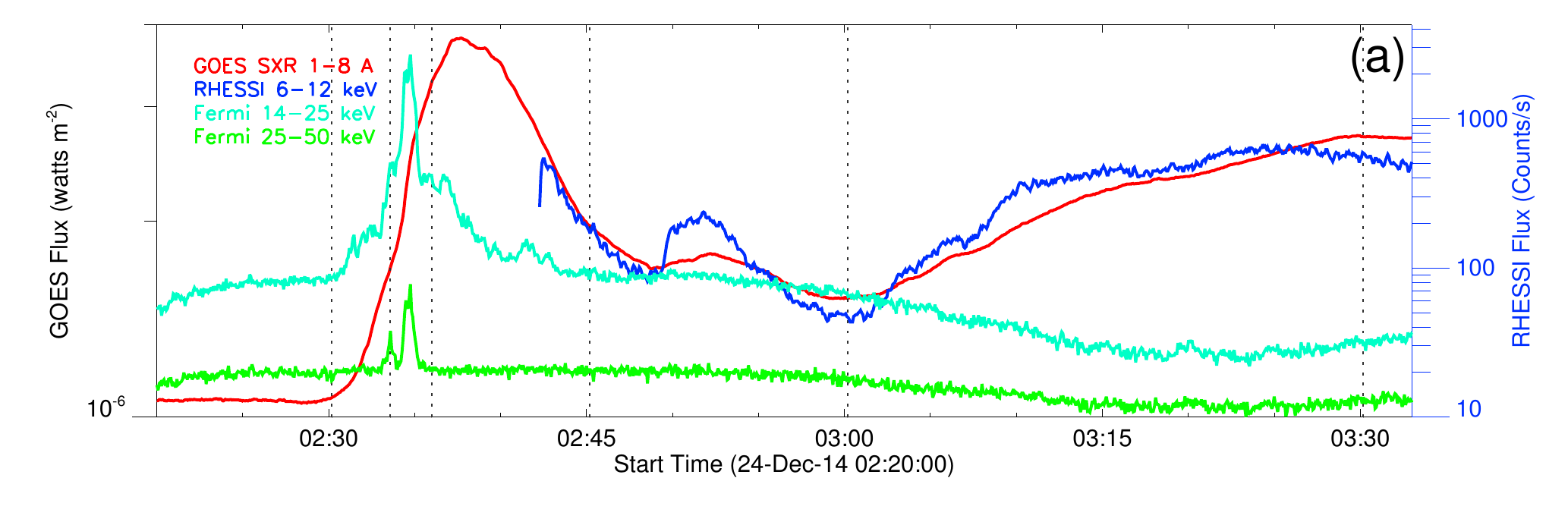}\vspace{-0.03\textwidth}}
\center {\includegraphics[width=15cm]{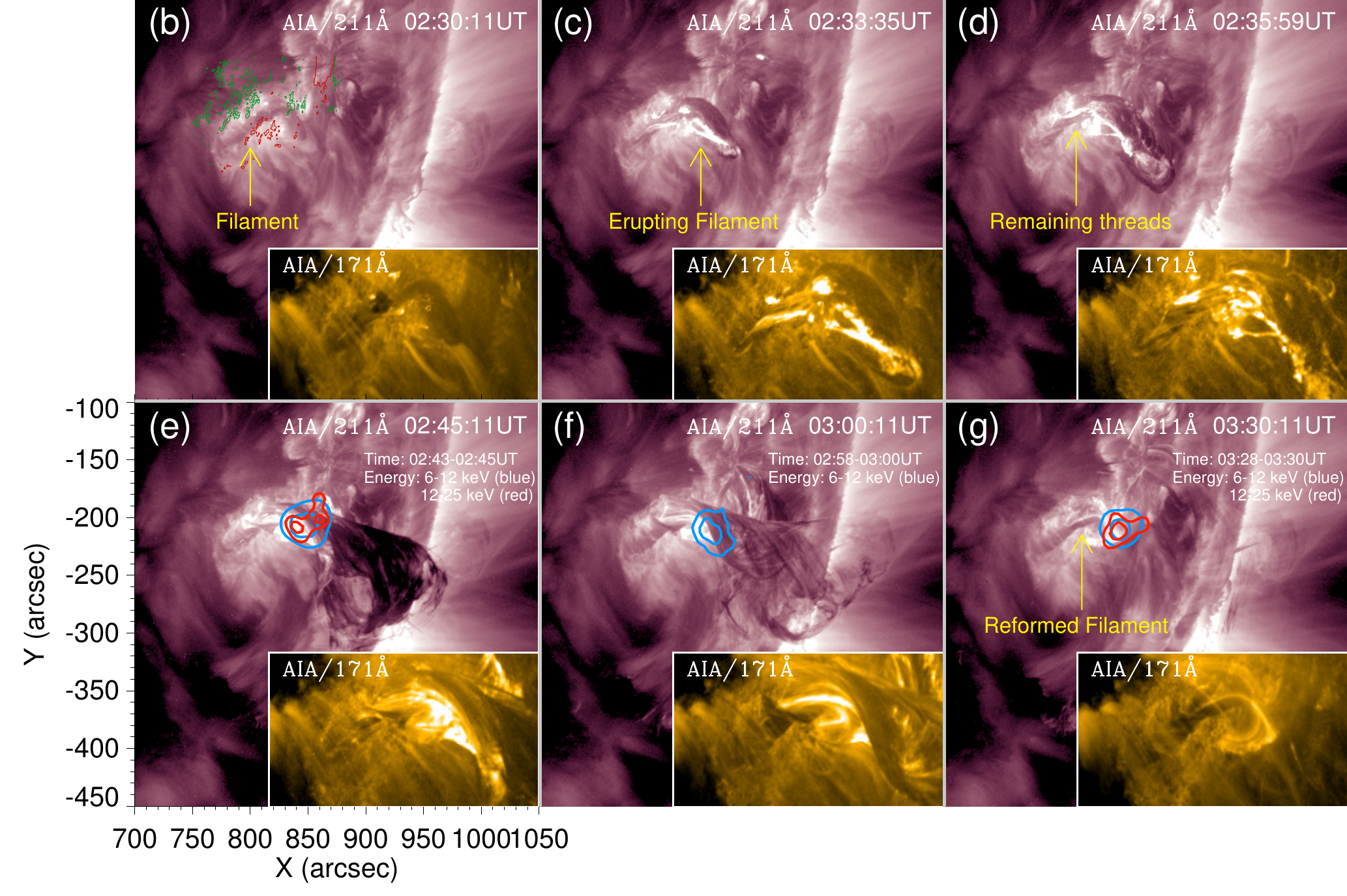}\hspace{0.065\textwidth}}
\caption{(a) \textit{GOES} SXR 1--8 {\AA} (red) and \textsl{RHESSI} 6--12 keV (blue) fluxes showing the temporal evolution of the SOL2014-12-24T02:30UT flare. \textsl{Fermi} HXR 14--25~keV (cyan) and 25--50~keV (green) fluxes are also plotted. The six vertical dashed lines denote the times of the further panels. 
(b)--(g) Time sequence of AIA 211 {\AA} and 171 {\AA} images displaying the rising, writhing, partial eruption and confinement of the filament. The contours in green (red) in Panel (a) represent the negative (positive) polarity of the magnetic field. The contours in blue and red in Panels (e)--(g) denote \textsl{RHESSI} 6--12 and 12--25 keV sources, respectively. The detectors 3F--8F were selected for reconstructing the HXR sources.}
\label{1224_aia}
\end{figure*}

As an alternative explanation for a vertical splitting into erupting and remaining flux, \citet{liurui12_filament} suggested that the coronal configuration at the onset of some partial eruptions contains an already vertically split MFR, equivalent to a double-MFR equilibrium (a ``double-decker MFR''). This was suggested by the observation that a partially erupting filament was vertically split into two branches already for many hours before the eruption of the upper branch. The idea was supported numerically \citep{kliem14} and observationally \cite[e.g.,][]{cheng14_formation,zhu14}. \citet{liurui12_filament} also pointed out that a vertical split and subsequent partial eruption are possible in the much simpler configuration of an MFR above arcade-type flux, separated by an X-type magnetic structure referred to as a hyperbolic flux tube (HFT). In this case, the lower, surviving branch of the filament does not conform to the widespread assumption of a flux rope structure, but rather is contained in a magnetic arcade. At present it is unclear which of the three suggested configurations---MFR with a BP or an HFT, or a double-decker MFR---represent the typical magnetic configuration prior to partial eruptions.

Another category of partial eruptions is given by a horizontal splitting. It happens quite often that an eruption does not comprise a whole filament or filament channel, but only a section of it. The other section may remain stable or rise delayed within the same, typically complex eruption, or fully separately eruption. Prominent examples are the Halloween events in AR~10468 on 2003 October~28 and 29 \citep{zuccarello09}. The phenomenon also occurs during asymmetric filament eruptions \citep{tripathi-isobe06}, even on smaller scales, including minifilaments \citep{panesar17}. Such horizontally split partial eruptions arise naturally if a fully coherent MFR has not (yet) formed, but the source region is composed of MFR and arcade sections \citep{guo10_filament,guo13_qmap, chintzoglou15}. Alternatively, a horizontal splitting may occur if the tension force of the overlying flux varies sufficiently along a filament channel. Horizontal splits of current-carrying flux potentially possess a high practical relevance because they can produce a sequence of CMEs, of which the subsequent ones generally tend to reach a high velocity, catching up with and then interacting with previous CMEs in the interplanetary space \citep{gopal2001,liulijuan17}.

\begin{figure*}
\center {\hspace{-0.06\textwidth} \includegraphics[width=15cm]{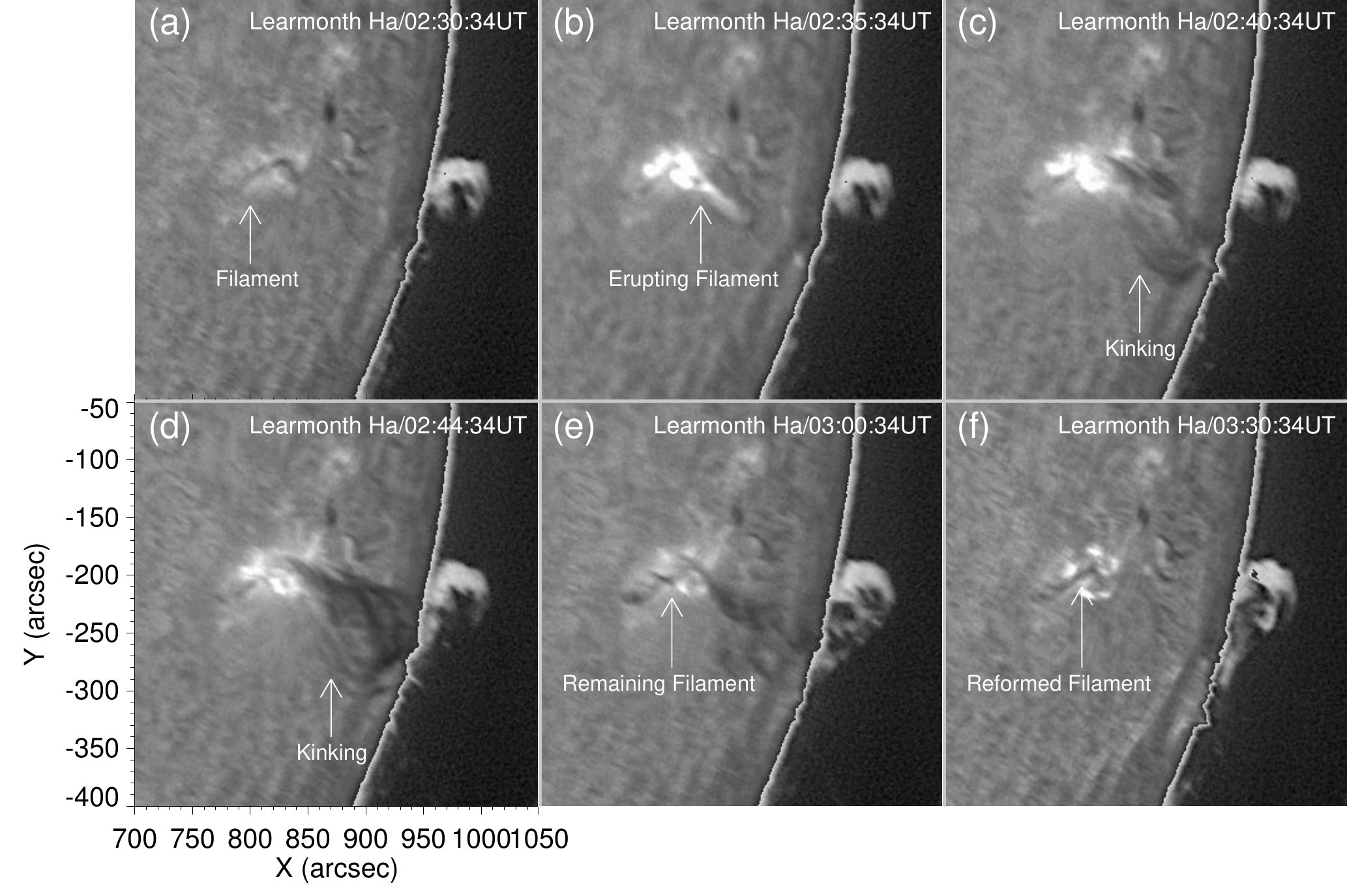}}
\caption{Time sequence of H$\alpha$ images showing the evolution of the failed partial filament eruption.}
\label{1224_ha}
\end{figure*}

In the present paper, we focus on the far more frequent phenomenon of partial eruptions that involve a vertical splitting of flux. We analyze two events which unambiguously show the internal reconnection implied by the BP-based mechanism for partial eruptions by \citet{gibson06_apjl}. This is made possible by the high resolution, high cadence, and multi-temperature sensitivity of the Atmospheric Imaging Assembly \citep[AIA;][]{lemen12} on board the \textsl{Solar Dynamics Observatory} \citep[\textsl{SDO};][]{pesnell12}. The first event is of interest also because it displays clear signatures of reconnection with ambient flux, leading to the nearly complete exchange of one footpoint, a partial exchange of the other footpoint, and a partial escape of filament material along presumably open field lines. In Section~\ref{s:Instruments}, we introduce the instruments. In Section~\ref{s:Observations}, we present observations and results, which are followed by a summary and discussion in Section~\ref{s:Summary}.

\section{Instruments}\label{s:Instruments}

We primarily use data from the AIA, which images the plasma in the corona through seven EUV and two UV passbands almost simultaneously with the response temperatures in the range of 0.06--20~MK. The cadence of EUV (UV) passbands is 12~s (24~s) and the pixel size is 0.6\arcsec. The Helioseismic and Magnetic Imager \citep[HMI;][]{schou12}, also on board \textsl{SDO}, provides magnetic field data with a cadence up to 45~s and a pixel size of 0.5\arcsec. Moreover, we make use of H$\alpha$ data acquired by Learmonth Solar Observatory and Cerro Tololo International Observatory which belong to the Global Oscillation Network Group (GONG). The H$\alpha$ data with a higher cadence (12 s) and higher resolution ($\sim$0.32\arcsec) from the New Vacuum Solar Telescope \citep[NVST;][]{liuzhong14} of the Yunnan Observatories are also utilized. In order to determine whether an eruption is successful, we take advantage of the Large Angle and Spectrometric Coronagraph \citep[LASCO;][]{brueckner95} white-light data. The \textsl{GOES}, \textsl{Reuven Ramaty High-Energy Solar Spectroscopic Imager} \citep[\textsl{RHESSI};][]{linrp02}, and \textsl{Fermi} satellites provide the soft X-ray (SXR) and hard X-ray (HXR) fluxes of the associated flares, respectively. In addition, we also explore the radio spectrograph data observed at Learmonth and the WAVES instrument \citep{Bougeret95} on board \textsl{WIND}. 

\section{Observations and Results}\label{s:Observations}
 
\subsection{A Confined Partial Filament Eruption on 2014 December 24}\label{ss:confined}

\begin{figure*}
\center {\hspace{-0.06\textwidth} \includegraphics[width=15cm]{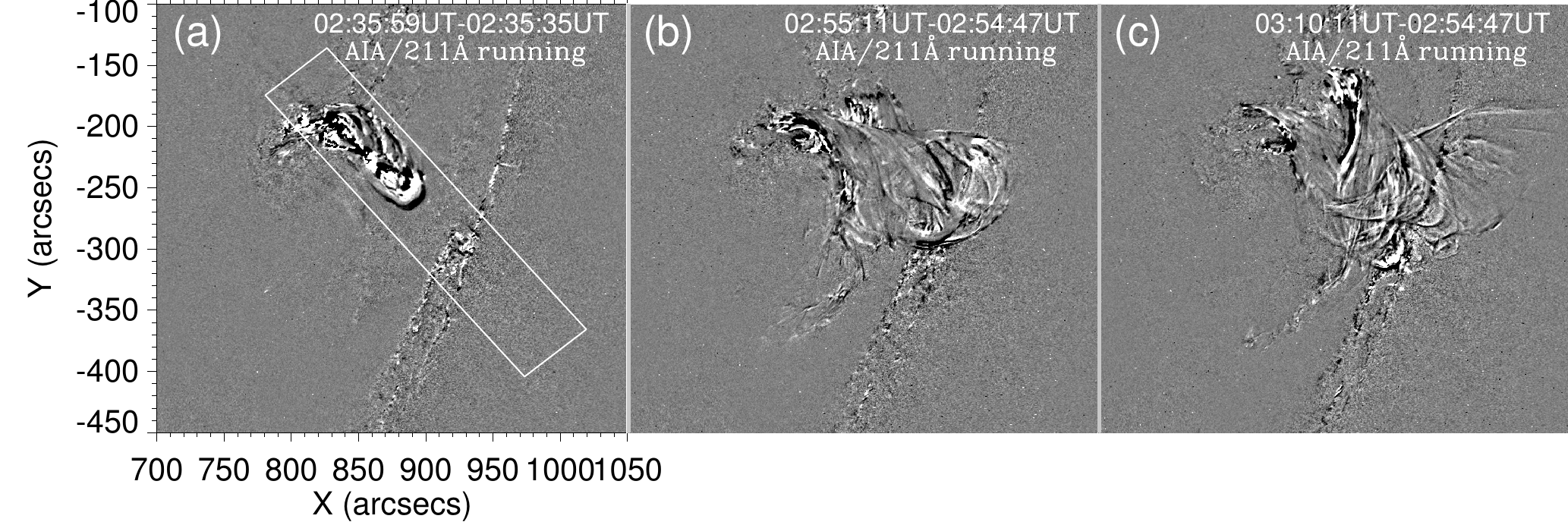}\vspace{-0.01\textwidth}}
\center {\hspace{ 0.00\textwidth} \includegraphics[width=16.2cm]{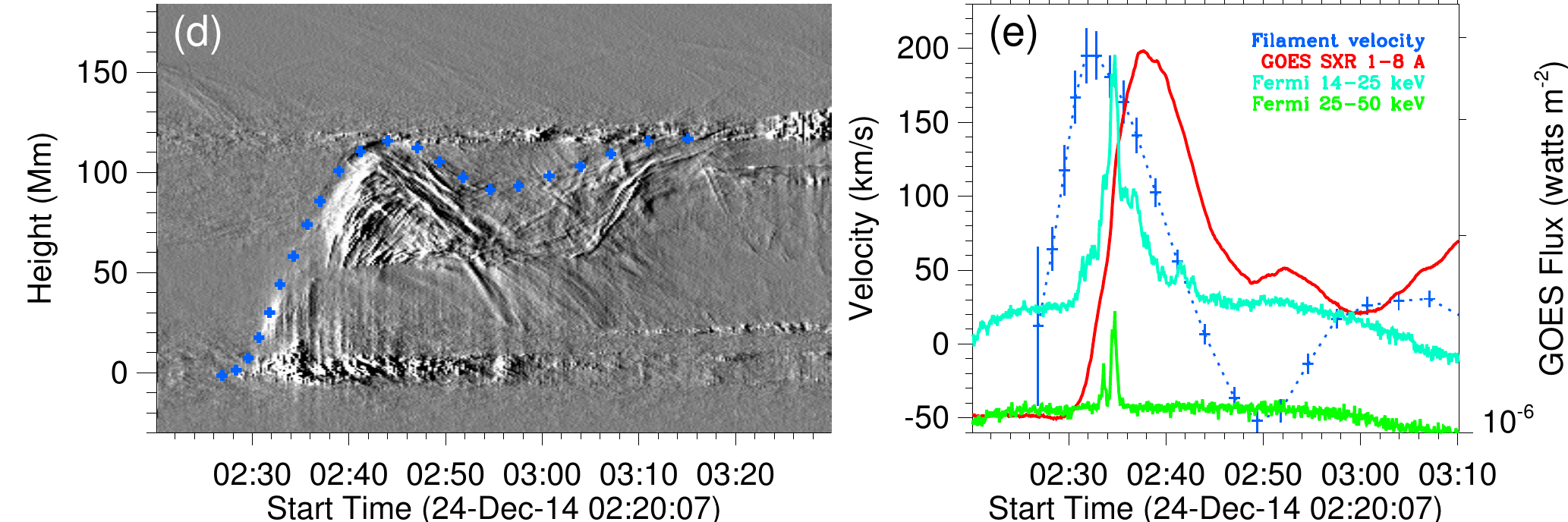}}
\caption{(a)--(c) AIA 211 {\AA} running difference images showing the writhing and confinement of the filament. The oblique rectangle in Panel (a) denotes a slice along the direction of the eruption. (d) The slice-time plot with the blue pluses overlaid denoting the height-time measurements of the eruption. (e) Temporal evolution of eruption velocity (blue), \textsl{GOES} SXR 1--8 {\AA} flux (red), and \textsl{Fermi} 14--25 and 25--50 keV fluxes (cyan and green, respectively).}
\label{1224_ht}
\end{figure*}

On 2014 December 24, a filament in NOAA active region 12241 near the solar west limb erupts partially and causes a \textsl{GOES} C3.7 class flare which begins at $\sim$2:30~UT and peaks at $\sim$2:37~UT. Figure~\ref{1224_aia}(a) shows the temporal profile of the \textsl{GOES} SXR 1--8 {\AA} flux, which indicates that the flare has two energy release stages. The first one also leads to significant particle acceleration as evidenced by the appearance of an HXR burst in the \textsl{Fermi} 14--25 and 25--50~keV bands. The second stage starts at 3:00~UT and corresponds to a more gradual increase of the SXR 1--8~{\AA} and HXR 6--12~keV fluxes, which lasts for more than 30 minutes but does not include strong or impulsive HXR emission at energies above 25~keV. 

\subsubsection{Filament Splitting, Presumably Caused by the Helical Kink Instability}

Prior to the eruption, the filament lies along the PIL of the active region and its main part takes on a forward-S shape with the middle section roughly oriented from northwest to southeast along the section of the PIL where flux cancellation has been prominent during the preceding days. The curved northwest filament section has a short extension further to the northwest, which is nearly straight in projection and extends to the southern edge of the strong, leading sunspot of the active region. The photospheric polarities in the end regions of the filament imply a sinistral chirality \cite[according to the definition of filament chirality in][]{martin98} and right-handed helicity for a force-free equilibrium. 

At $\sim$2:20~UT, the northwest curved section and part of the middle straight section of the filament start to rise up slowly. This changes suddenly to a much faster rise at $\sim$2:28~UT (Figures~\ref{1224_aia}(b)--(d) and \ref{1224_ha}(a)--(c)). With a small delay, significant plasma heating, manifested as EUV brightenings, takes place under the rising section (Figures~\ref{1224_aia}(c) and \ref{1224_ha}(b)). Some brightened threads are also seen to be intertwined with the dark threads in the remaining southeast part of the filament in a right-handed sense (Figure~\ref{1224_aia}(c)--(d)). This pattern becomes visible with the first brightenings of the associated flare at $\sim$2:30~UT, which suggests that an MFR exists already at the onset of the eruption.

The filament axis significantly changes its direction during the rise. The top part of the axis rotates clockwise by at least $90^\circ$ to a nearly south-north orientation. As a result, in the period 2:30--2:40~UT, the filament axis writhes strongly (Figures~\ref{1224_aia}(b)--(e), \ref{1224_ha}(a)--(d) and associated animations). This characteristic suggests that the filament may experience the helical kink instability by which part of its twist is transformed into writhe. However, this conclusion cannot be drawn unambiguously, because an axis rotation of about $90^\circ$ can also be produced by the torus instability if the ambient field has a significant component along the filament axis \citep{kliem12}. Here it is not known whether the latter condition is fulfilled and how much the true rotation angle about the direction of ascent differs from the rotation angle seen in the plane of sky of AIA. The confined nature of the eruption provides an additional indication for the helical kink as the initiating instability, which is not conclusive. The helical kink always saturates quickly, while the torus instability requires a strong overlying flux at the terminal height of the eruption in order to stop there. Overall, the data suggest the occurrence of the kink instability but do not prove it. 

\begin{figure*}
\center {\hspace{0.0\textwidth} \includegraphics[width=15cm]{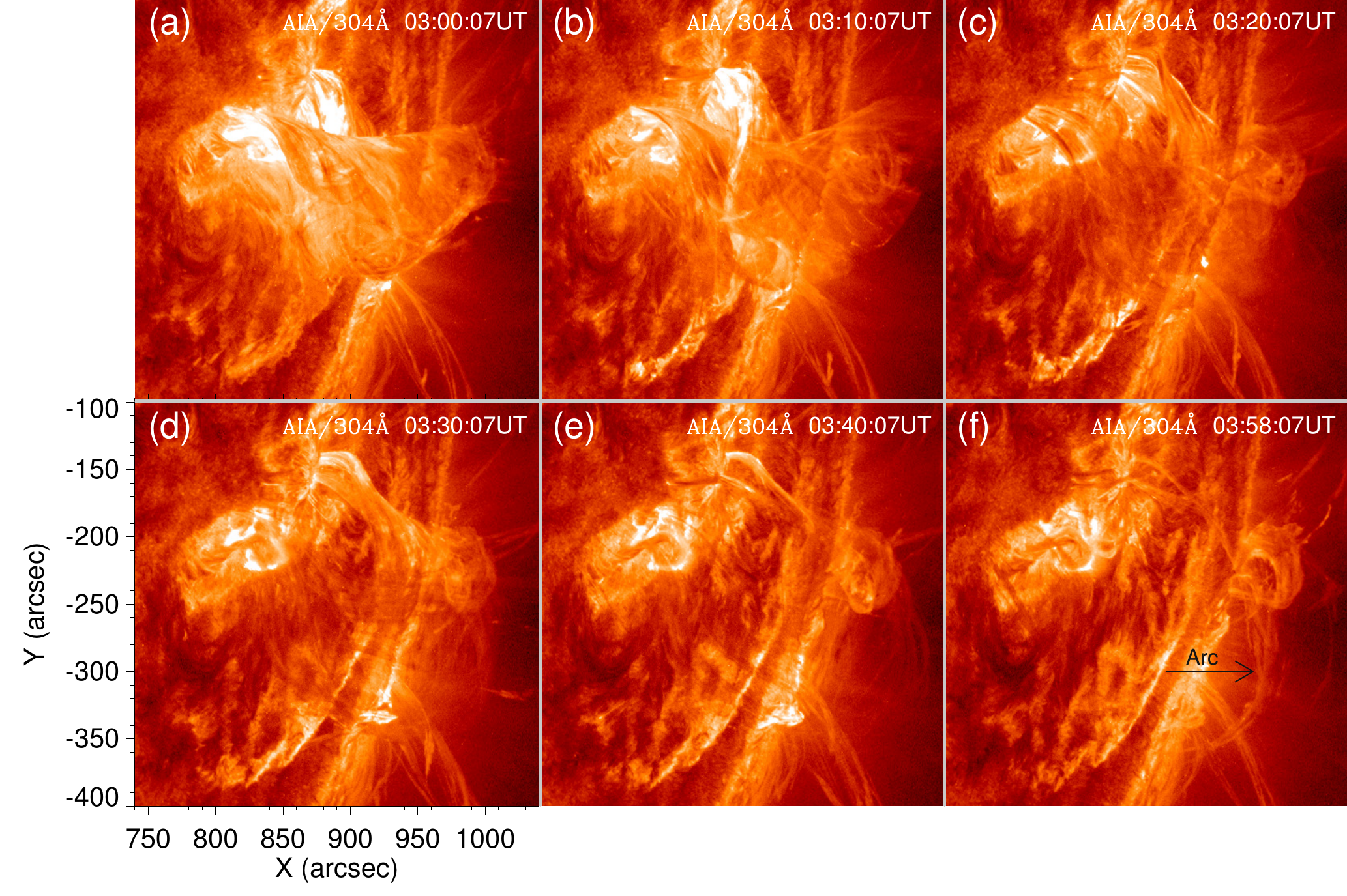}} 
\caption{Time sequence of AIA 304~{\AA} images showing during the second energy release stage: (a)--(d) the northwestward progression of the eruption's positive footpoint into the area of the leading sunspot and the forming connections to open flux; (e)--(f) the arc connecting to the negative-polarity network flux $10^\circ$ south of the filament (marked by an arrow) and formation of a weak remote flare ribbon.}
This figure is available online as an animation that displays the partial eruption and reconnection with the ambient field of the filament.
\label{1224_arc} 
\end{figure*}

Only a part of the filament erupts. In Figures~\ref{1224_aia}(d)--(f) and \ref{1224_ha}(c)--(e), one can clearly see that the southeast part, including the small curved end section and some of the middle, straight section, remains in the original place throughout the eruption. Furthermore, under the erupting northwest part of the filament, a bundle of threads remains at the original position. This material completes the post-eruption filament, which has the same position as the original one with a slightly shorter northwest curved section and no further northwestward extension (Figures~\ref{1224_aia}(f)--(g) and \ref{1224_ha}(e)--(f)). These observations clearly indicate that the unstable northwest part of the filament structure splits vertically in two parts (Figure~\ref{1224_aia}(c)--(d)). In this event, the split occurs below most of the visible filament material.

\begin{figure*}
\center {\hspace{-0.06\textwidth}
\includegraphics[width=15cm]{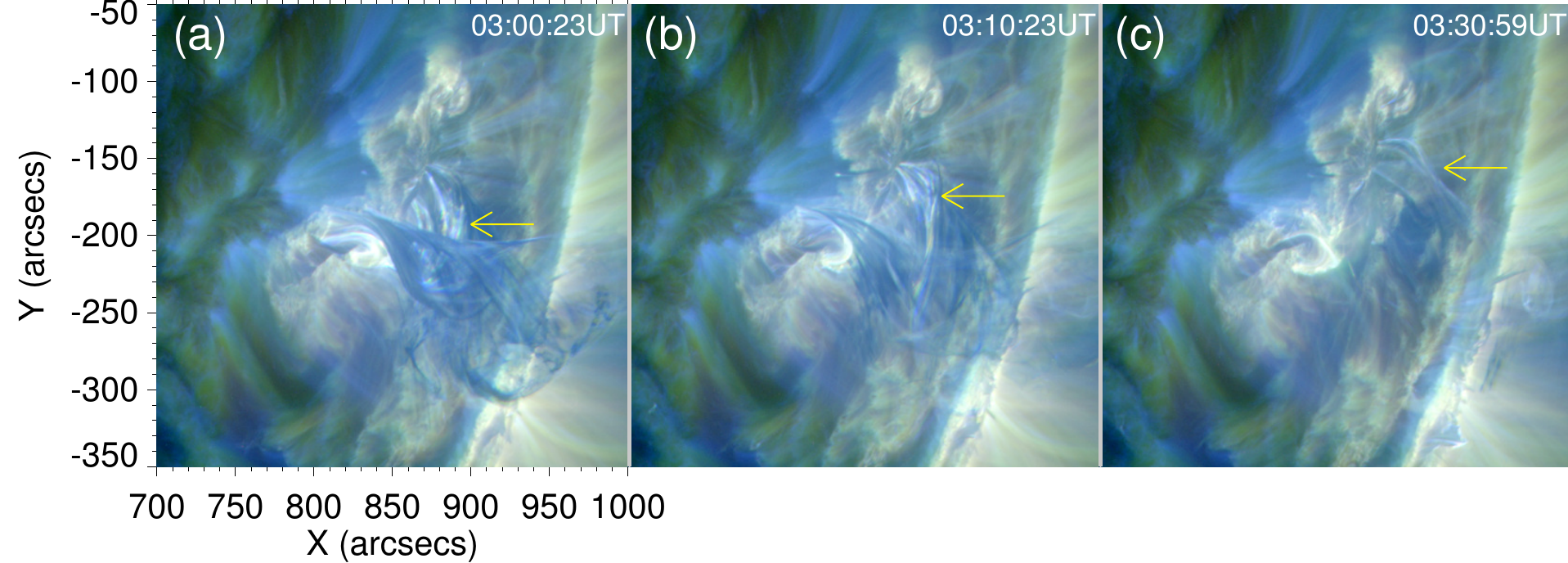}
\includegraphics[width=15cm]{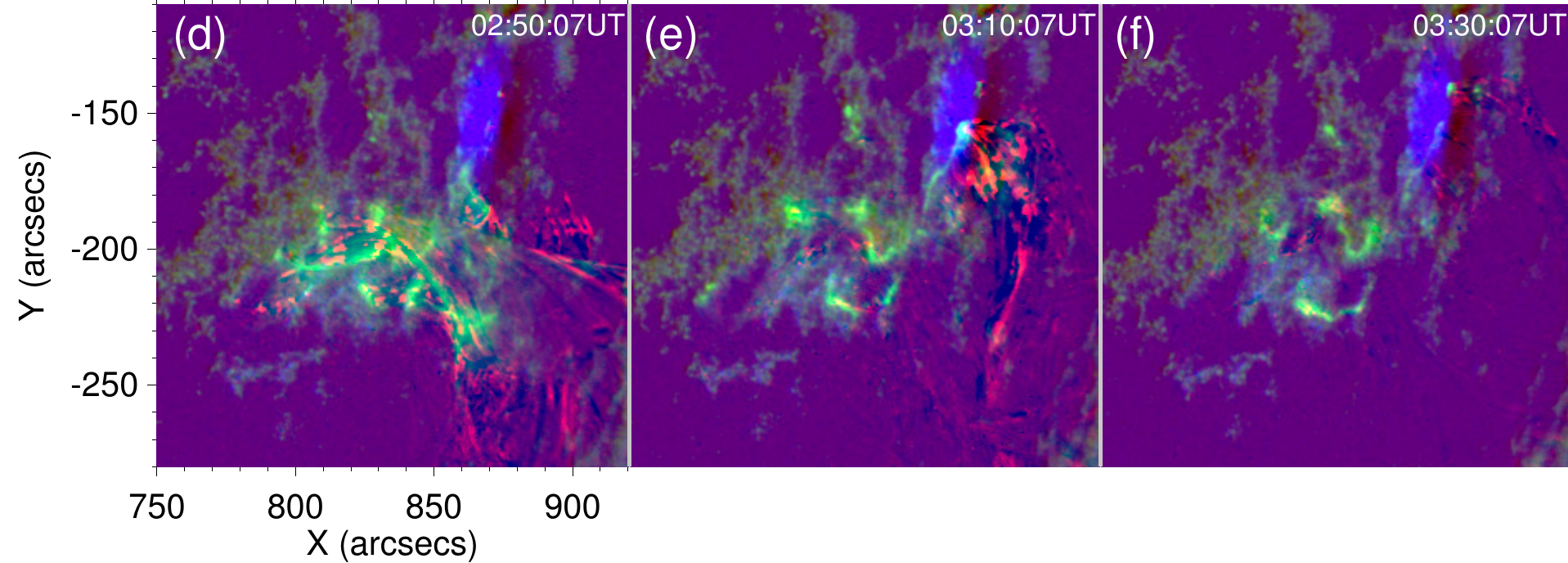}\hspace{0.05\textwidth}}
\caption{(a)--(c) Composite images of the AIA 211 {\AA}, 193 {\AA}, and 171 {\AA} passbands showing the heating of the filament material that is trapped in the newly formed connection to the sunspot (marked by the arrows). (d)--(f) Composite images of the AIA 304 {\AA}, 1600 {\AA} passbands, and HMI line-of-sight magnetograms displaying the rise of the newly formed structures (arrows), the draining of the material trapped in them, the evolution of the flare ribbons, and the distribution of the photospheric magnetic flux.}
This figure is available online as two animations that detailedly display the partial eruption of the filament and subsequent reconnection.
\label{1224_foot}
\end{figure*}

Correspondingly, EUV brightenings appear between the erupting and surviving filament segments (Figures~\ref{1224_aia}(c)--(d) and \ref{1224_ha}(b)--(c)). \textit{RHESSI} HXR sources in the energy ranges of 6--12 and 12--25 keV appear almost coincident with the strongest EUV brightenings (Figure~\ref{1224_aia}(e)--(g)). (Note that \textit{RHESSI} is in spacecraft night in the early part of the event.) The EUV brightenings evolve into cusp-shaped flare loops, as visible in the AIA 211 {\AA} and 171 {\AA} passbands. They are located above the surviving filament, which is most clearly seen in Figure~\ref{1224_aia}(f) and (g). In the chromosphere, the two ribbons formed along the footpoints of these loops are situated at both sides of the surviving filament (Figure~\ref{1224_ha}(d)--(f)). All of these features indicate that magnetic reconnection occurs in a vertical current sheet that forms between the erupting and surviving parts of the filament. 

Prior to the eruption, the filament does not show any indication of a split in two branches (Figure \ref{1224_aia}(b)). During the eruption, some of the brightened threads in the rising northwest part are seen to continue into the remaining southeast part with a common twisting pattern (Figure \ref{1224_aia}(c)--(d)). This suggests that the rising and remaining parts of the filament belong to the same MFR prior to the eruption, so that the reconnection associated with the vertical splitting is internal reconnection in agreement with the model by \citet{gibson06_apjl}. 

We cannot check for the existence of a BP at the time of the eruption because the active region lies then close to the limb, where the horizontal field component at the photospheric PIL is prone to substantial measurement uncertainty. Testing for the existence of BPs at an earlier time when the active region was within 45~deg from central meridian, is not reliable either, because the filament showed another eruption near this longitude.

\begin{figure}
\center {\vspace{0.04\textwidth} \includegraphics[width=7cm]{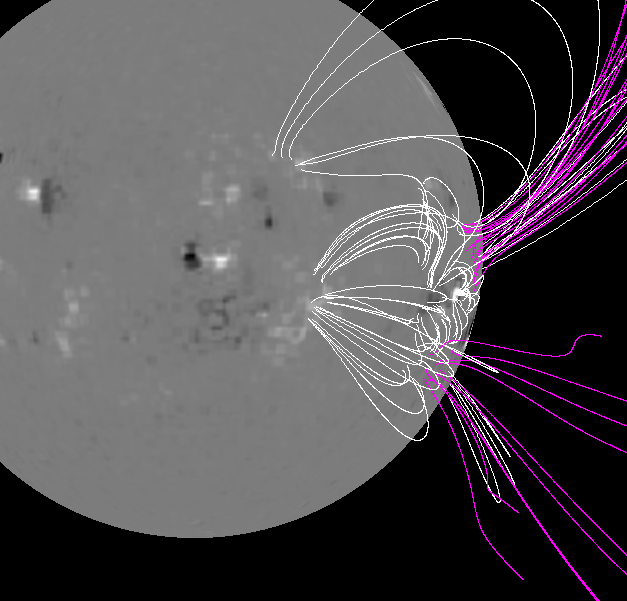}} 
\caption{PFSS magnetic field structures above the NOAA active region 12241. The white and pink curves show closed and open magnetic field lines, respectively.}
\label{pfss} 
\end{figure}

\subsubsection{Kinematics and Confinement of the Eruption} 

The low coronal features such as the erupting filament, EUV brightenings, HXR sources, cusp-shaped loops, and two flare ribbons conform to the scenario of a successful filament eruption which evolves into a CME. Nevertheless, through carefully inspecting the AIA and LASCO/C2 images, we find that this partial filament eruption is actually confined in the corona and does not generate a well-defined CME. Only a portion of the erupted material escapes into the background solar wind. As inferred in Section~\ref{ss:reconnection} and similarly by \citet{zhengrs17}, the escape is probably due to reconnection between the erupted filament flux and nearby open flux. From the AIA 211~{\AA} running difference images in Figure~\ref{1224_ht}(a)--(c), one can see that the rise of the filament eventually stops. 

In order to study the kinematics of the confined partial filament eruption, we put a slice along the direction of the eruption and make the slice-time plot shown in Figure \ref{1224_ht}(d). For the sake of enhancing the contrast of the plot, a 60{\arcsec} wide slice is selected and then averaged. The result shows that the partial eruption primarily experiences an accelerated fast rise, followed by a deceleration to zero velocity and even a small retreat. Subsequently, some threads at the upper edge of the filament turn to a very slow rise, while other threads experience a second phase of considerable acceleration from about half the top height to the front (during $\sim$3:00--3:15~UT). At the same time, even other structures show a downward shift. In this latter phase, the filament has obviously lost its coherence and its threads show a mix of several effects. Some in the middle are moderately accelerated upwards, the front edge rises very slowly with negligible acceleration, other threads in the middle continue to writhe, which appears as a slight downward motion in the given projection, and most threads already drain to the surface. The structures that remain at the front edge after the draining form a faint arc (Figure~\ref{1224_arc}) which continues to rise very slowly until it ceases at $\sim$4:35~UT. This small velocity, the absence of a coronal cavity above the eruption site in the EUV images, and the absence of a well defined CME signature in the LASCO data clearly indicate that this eruption remains confined.

The height-time measurements of the front edge are displayed as the blue pluses in Figure~\ref{1224_ht}(d). Applying the first-order numerical derivative, the velocity as a function of time is derived as shown in Figure~\ref{1224_ht}(e). One can see that the velocity of the partial eruption quickly increases to $\sim$200 km s$^{-1}$ in about 5 minutes, with an acceleration of $\sim$670 m s$^{-2}$. Subsequently, the velocity quickly decreases to zero at $\sim$2:44~UT and changes sign to $-50$ km s$^{-1}$ at $\sim$2:50~UT, which must result from the downward tension force of the overlying field. The erupting filament stretches the overlying field, which in return produces this force to counteract the eruption. The second significant acceleration of threads in the middle of the erupted filament coincides with the onset of the second energy release of the flare and may thus be related to the topology changes resulting from the underlying reconnection. We discuss this in Section~\ref{ss:reconnection}. 

Figure \ref{1224_ht}(e) shows that the fast rise of the partially erupting filament commences suddenly and earlier than the flare. Specifically, the onset of the filament acceleration precedes that of the flare SXR 1--8 {\AA} emission by $\sim$4 minutes; the peak filament velocity also appears earlier than the SXR peak by $\sim$5 minutes. These delays cannot be explained by occultation, because the eruption is seen nearly in a side view. Based on the fact that the flare emission is closely related to the energy release by reconnection, the delayed onset of reconnection strongly supports the view that the eruption is initiated by a loss of equilibrium (ideal MHD instability) of a part of the filament rather than by the reconnection \citep[for a similar conclusion for another event, see][]{song15_apjl}. A loss of equilibrium is also suggested by the sudden onset. Furthermore, the timing of the two onsets is consistent with the model for partial eruptions by \citet{gibson06_apjl} because there is no seed for prompt flare current sheet formation in the BPS topology. Rather, the upper, unstable part of the MFR must rise by some height interval until the splitting creates a flare current sheet within the volume of the MFR. Only then the fast flare reconnection can set in. 

\subsubsection{Reconnection With the Ambient Field}\label{ss:reconnection}

The second enhancement of the SXR emission from 3:00 UT (Figure \ref{1224_aia}(a)) indicates that there is a continuous, gradually amplifying energy release, although the front of the eruption shows only a further slow rise without significant acceleration (Figure \ref{1224_ht}(d)). The top part of the filament continues to writhe, but there are hardly any strong EUV brightenings which could indicate energetically significant reconnection in this area. However, new connections between the erupted filament and the main sunspot northwestward of the original filament form from $\sim$02:37~UT onward and the filament material trapped and draining in them brightens strongly throughout the second SXR enhancement. Figure~\ref{1224_foot}(a)--(c) displays composite images of the AIA 211 {\AA}, 193 {\AA}, and 171 {\AA} passbands which show the new structures. They form in or near the area of reconnection between the erupted and underlying remaining filament parts, brighten considerably, and subsequently show a relatively fast rise to the top of the erupted filament. This rise is the most prominent feature during $\sim$3:00--3:15~UT (Figure~\ref{1224_ht}(d)), but continues until $\sim$3:40~UT outside the area selected by the slit in Figure~\ref{1224_ht}(a). The footpoints of the new structures lie in the main sunspot, remote from the original end of the filament. This implies reconnection of the erupting flux with ambient flux rooted in the sunspot. Figure~\ref{1224_foot}(d)--(f) displays composite images of the AIA 304 {\AA} and 1600 {\AA} passbands, as well as HMI line-of-sight magnetograms. In addition to the brightening and rise of the new structures, it is showed that the flare ribbons extend northwestward from the filament along the PIL between the sunspot and the dispersed main negative polarity of the active region. This indicates that further reconnection is triggered under the rising new structures. Indeed, a second arcade of flare loops forms at this section of the PIL, starting at $\sim$3:11~UT in the hottest (131~{\AA}) AIA channel (not shown here). The timing of all these features suggests that the second flare enhancement after 3:00~UT is due to the formation and rise of the new structures. 

When the new structures arrive at the front edge of the eruption, much but not all of their absorbing material has drained. The remaining material outlines a faint large arc which extends from the sunspot to dispersed negative network $\sim$10$^\circ$ south of the filament (Figure~\ref{1224_arc}(f)). Since both footpoints of the arc lie in areas that originally did not show any connection to the filament, the flux holding the arc must have reconnected twice. Before the arc forms, material drains from the top part of the erupted filament to the arc's southern footpoint area and several brightenings occur there subsequently, including a weak but extended flare ribbon (Figures~\ref{1224_arc} and \ref{1224_foot}); so it is clear that ambient flux rooted there becomes involved in the eruption. 

\begin{figure}
\center {\hspace{-0.06\textwidth} \includegraphics[width=8cm]{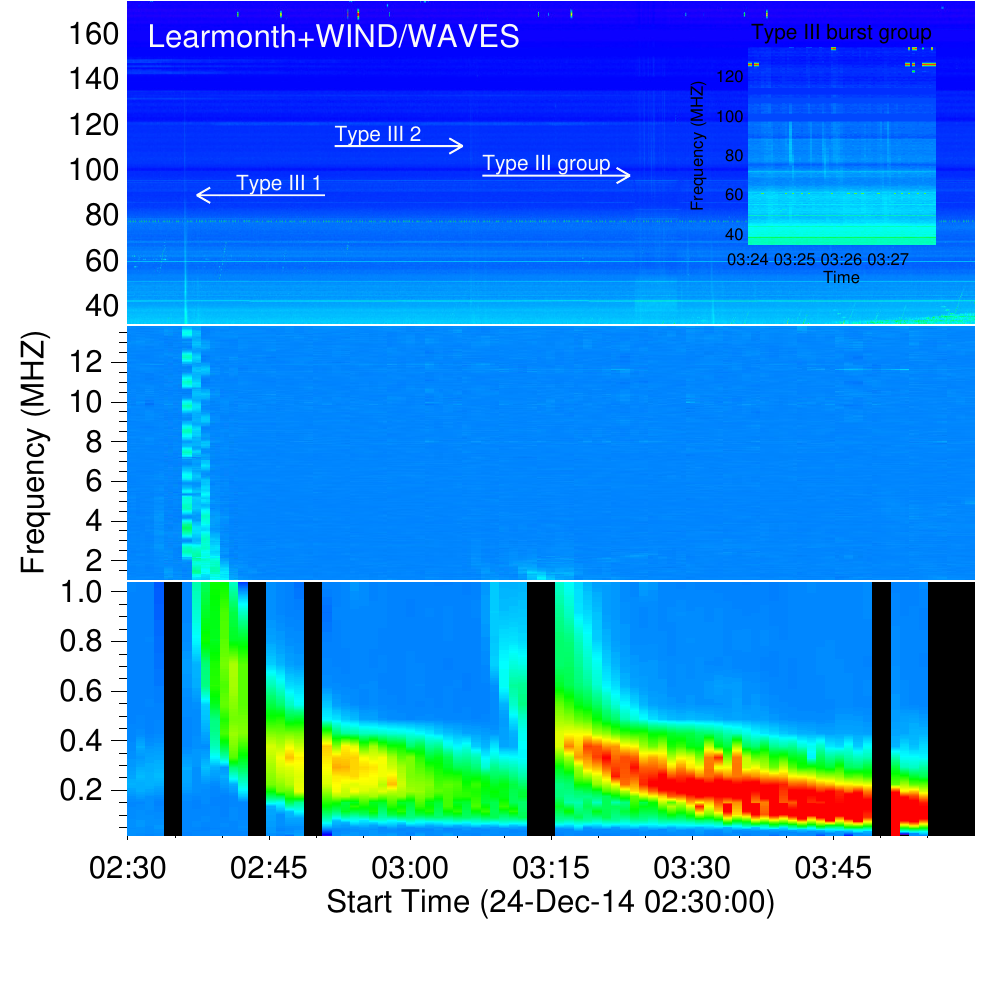}}
\caption{Learmonth and \textsl{WIND}/WAVES radio spectrogram showing the appearance of two type III radio bursts (type III burst 1 and type III burst 2) and a type III radio burst group (zoomed display in the inset).}
\label{1224_radio}
\end{figure}

The EUV images additionally indicate that part of the erupted filament flux reconnects with ambient open flux. Between $\sim$2:58 and $\sim$3:15~UT, some threads in the top part of the erupted flux turn nearly radial, suggesting that they have formed a connection with high-reaching, possibly open field lines, along which some blobs of filament material move outward. This is best seen in the cooler AIA 304~{\AA} channel and may contribute to the occasional appearance of cold, dense plasma in the solar wind \citep{zhengrs17}. A potential-field source-surface extrapolation (Figure~\ref{pfss}) suggests that, near the source region of the eruption, open flux is rooted only in the dispersed negative network south of the filament. Therefore, the flux from the negative footpoint of the erupting filament must undergo interchange reconnection with the open part of that network flux. This is confirmed by the shape of the opening filament threads (Figure~\ref{1224_foot}(b)). The connection of the faint arc to the network area shows the other reconnection product. Subsequently, the relatively faint, radial structures sweep southward, accompanied by intermittent brightenings in different places, until they fade at $\sim$3:49~UT. Since filament material is trapped in the nearly radial flux, the southward displacement must be a true motion of the flux, not an apparent motion mimicked by a progressing illumination of a static structure. The southward displacement coincides with the occurrence of a small set of flare loops in the neighboring Active Region 12242. From the temporal coincidence, it appears possible that the eruption triggered some large-scale effects, although it was only partial and confined, and associated with only a moderate flare.

Interchange reconnection between filament flux and ambient flux is also indicated by the appearance of type III radio bursts. Figure~\ref{1224_radio} shows that a first type III burst starts at $\sim$2:37 UT with a start frequency of $\sim$130 MHz, which corresponds to a density of 2$\times$10$^8$ cm$^{-3}$, and quickly drifts to a frequency $\sim10^3$ times lower, where the ambient density is smaller by a factor $\sim10^6$. Thus, energetic electrons accelerated by reconnection in the flare escape to the interplanetary space. The temporal coincidence with the first signs of reconnection with ambient flux is remarkable and suggests that some open flux, rooted near the filament, was involved in this reconnection even before the connections to open flux were indicated by the EUV images. At $\sim$3:05~UT, shortly after the second stage of the flare begins and during the radial alignment of some filament threads, another type III burst appears with the same start and end frequencies. Although the underlying electrons may propagate along different field lines, this type~III burst supports the conjecture \citep{zhengrs17} that the radial filament threads are trapped on open flux. About 10 minutes later, a type III radio burst group with similar start frequencies appears and lasts for $\sim$15 minutes (see the inset in Figure~\ref{1224_radio}). This group of electron beams is generated when the connections between the erupted filament flux and ambient open flux show the rapid southward sweeping. Magnetic reconnection is likely to occur during such a strong change of magnetic flux and is indicated by the intermittent brightenings of the trapped material.

\subsection{A Successful Partial Filament Eruption on 2015 November 4}\label{ss:successful}

\begin{figure*}
\center {\includegraphics[width=16.9cm]{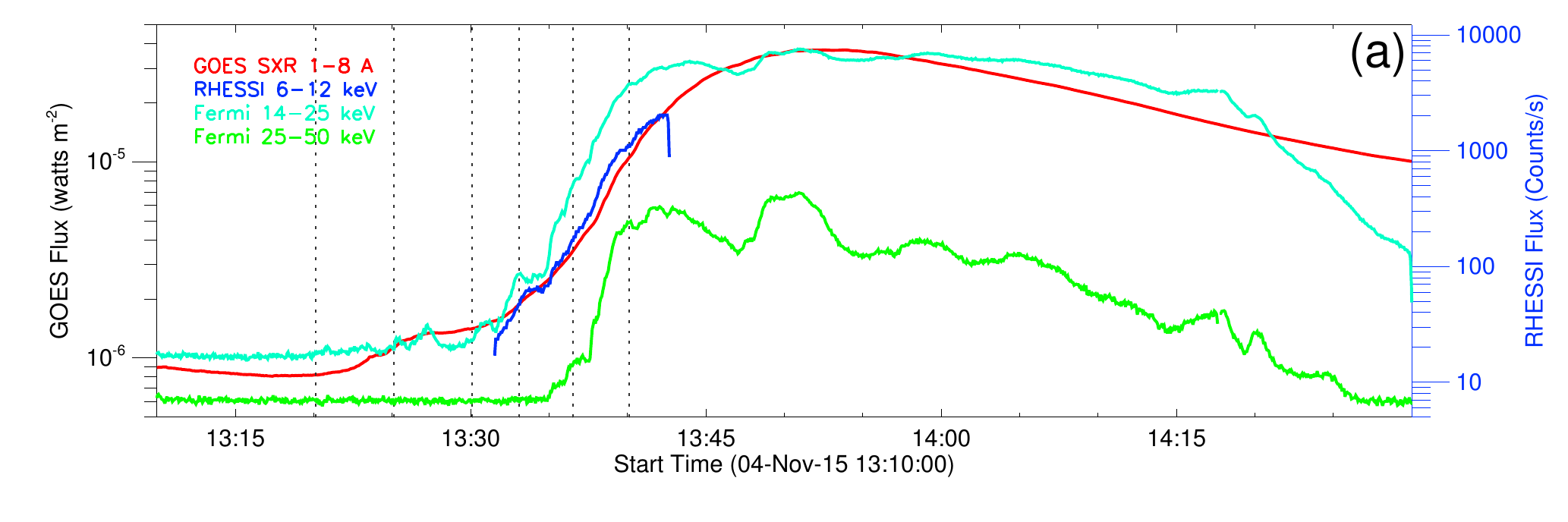}\vspace{-0.03\textwidth}}
\center {\includegraphics[width=15cm]{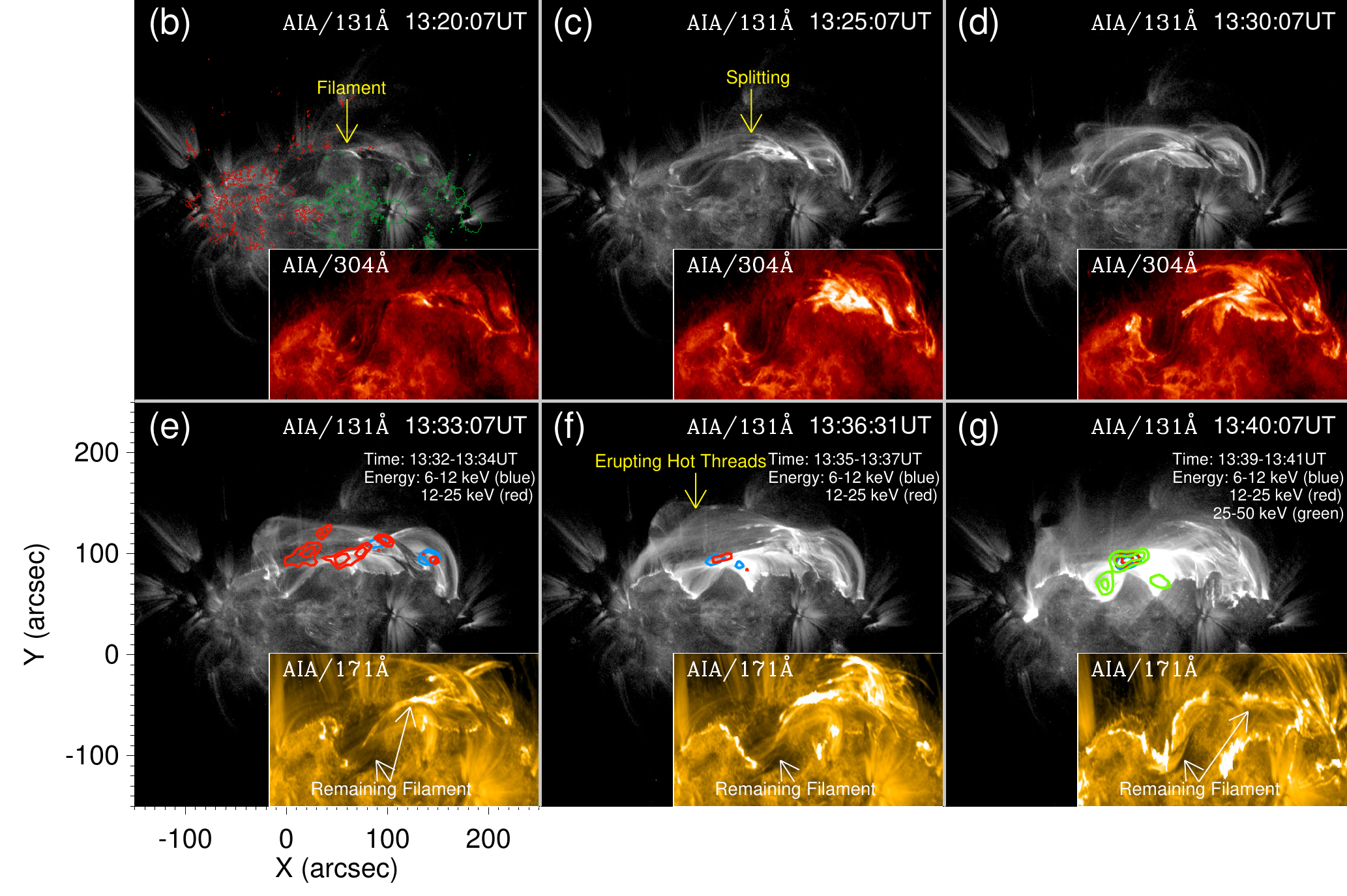}\hspace{0.065\textwidth}}
\caption{(a) \textsl{GOES} SXR 1--8 {\AA} (red) and \textsl{RHESSI} 6--12 keV (blue) fluxes showing the temporal evolution of the SOL2015-11-04T13:31UT flare. \textsl{Fermi} HXR 14--25~keV (cyan) and 25--50~keV (green) fluxes are also plotted. The six vertical dashed lines indicate the times of the further panels. 
(b)--(g) Time sequence of AIA 131 {\AA}, 304 {\AA}, 171 {\AA} images displaying the rise, splitting, and eruption of the filament. The contours in green (red) in Panel (a) represent the negative (positive) polarity of the magnetic field. The contours in blue, red, and green in Panels (e)--(g) denote \textsl{RHESSI} 6--12, 12--25, and 25--50 keV sources, respectively. The detectors 3F--8F were selected for reconstructing the HXR sources.}
This figure is available online as an animation that detailedly displays the rise, splitting, and partial eruption of the filament.
\label{1104_aia}
\end{figure*}

On 2015 November 4, a partial filament eruption takes place near the solar disk center and produces an M3.7-class flare and a halo CME. The start and end times of the flare in the NOAA reports are $\sim$13:31~UT and $\sim$14:13~UT, respectively. The duration of 42~min indicates that the flare is a long-duration event. Figure \ref{1104_aia}(a) shows the temporal profiles of the \textsl{GOES} SXR 1--8~{\AA}, \textsl{RHESSI} 6--12~keV, and \textsl{Fermi} 14--25 and 25--50~keV fluxes. One can see that the HXR emission also lasts for a long time. From the beginning at $\sim$13:31~UT, the \textit{Fermi} 14--25 and 25--50~keV fluxes continuously increase until $\sim$13:50~UT. After that, they start to decrease gradually and reach the background level at $\sim$14:30~UT.

\subsubsection{Filament Splitting} %Caused by Internal Reconnection}

\begin{figure*}
\center {\includegraphics[width=15cm]{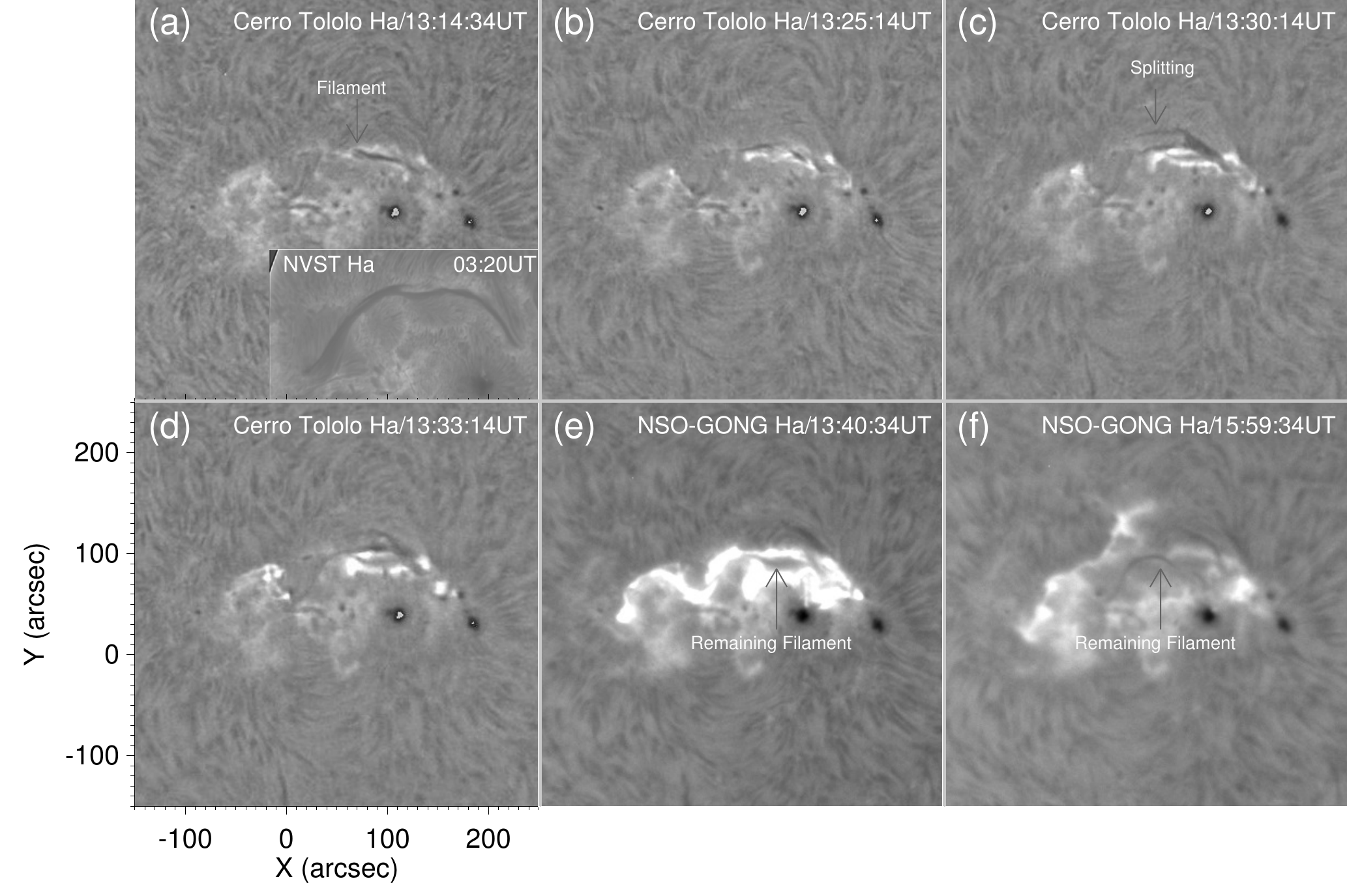}}
\caption{Time sequence of H$\alpha$ images showing the evolution of the partial filament eruption.}
\label{1104_ha}
\end{figure*}

\begin{figure*}
\center {\includegraphics[width=15cm]{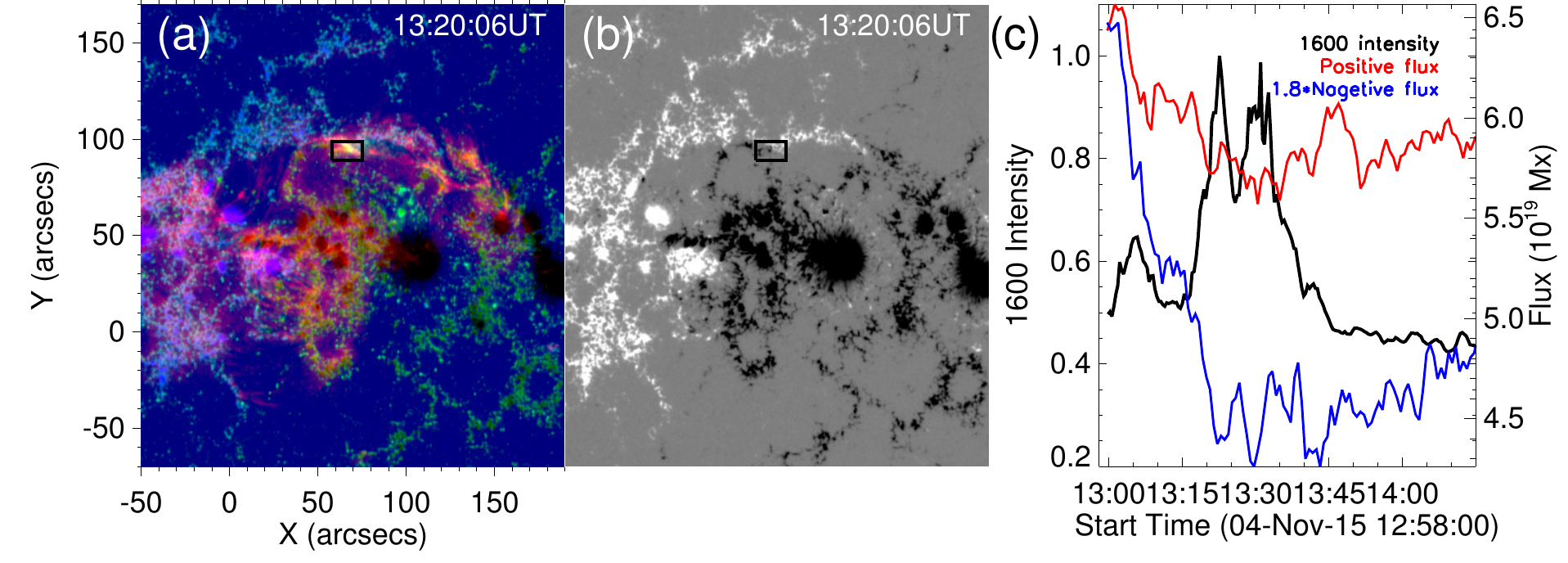}}
\caption{(a) Composite image of the AIA 304 {\AA}, 1600 {\AA} passbands, and HMI line-of-sight magnetogram displaying the locations of the initial EUV/UV brightening and magnetic cancellation. (b) HMI line-of-sight magnetogram; white (black) show positive (negative) polarity. (c) Temporal evolution of the integrated AIA 1600 {\AA} intensity (black), positive magnetic flux (red), and negative magnetic flux (blue) in the back box shown in Panels (a) and (b).}
This figure is available online as an animation that clearly displays the initial EUV/UV brightening and magnetic cancellation.
\label{1104_mag}
\end{figure*}

A filament channel exists along the PIL at the north side of the leading negative polarity of NOAA Active Region 12443 as a result of long-lasting converging flows and flux cancellation. Cool material appears along most of the filament channel and manifests as filament threads in the EUV and H$\alpha$ passbands (Figures~\ref{1104_aia}(b) and \ref{1104_ha}(a)). The right section of the filament is better visible, especially in the H$\alpha$ line, where the left section is clearly seen only in the high-resolution observations by the NVST (see the inset in Figure~\ref{1104_ha}(a)). The eruption starts with brightenings in the right section (Figure~\ref{1104_aia}(b)--(c)) at a location of strong flux cancellation (Figure~\ref{1104_mag}), but includes the flux along the whole extent of the filament from the beginning. After $\sim$13:32~UT, the flux rooted further eastward also becomes involved, which is associated with the onset of the impulsive flare phase. This spatial extension, studied in detail by \citet{wang17_nc}, is not relevant in the present context. 

The brightenings in the right section of the filament start at $\sim$13:15~UT in the EUV passbands and appear at various places along and between the threads in the body of the filament. During the following $\sim$8~min, they indicate a right-handed twisted (MFR) structure for the filament in all AIA channels which display cool, absorbing plasma (Figure~\ref{1104_aia}(b)). Simultaneously with the onset of the brightenings, the splitting begins by an initially slow rise of part of the threads, while the other threads remain at the original height. In this event, it appears that the split cuts through the body of the filament in this right section. 

The splitting is accompanied by a strong amplification of the EUV brightenings from $\sim$13:22~UT, preferentially between the separating parts (Figures~\ref{1104_aia}(c) and \ref{1104_ha}(b)--(c)). The brightenings clearly indicate internal reconnection within the filament-hosting MFR. Unfortunately, at their peak, they mask the evolution of the lower, remaining part of the filament. From $\sim$13:27~UT, a pair of flare ribbons begins to separate from the brightenings (Figures~\ref{1104_aia}(d) and \ref{1104_ha}(c)). Simultaneously, the arcade of flare loops begins to form at this location in the AIA 131~{\AA} images (Figure~\ref{1104_aia}(d)). As the brightenings between the splitting parts of the filament decrease in H$\alpha$ and most AIA passbands (except in the hottest ones), while the flare ribbons grow and amplify, the remaining dark filament material again becomes visible between the ribbons, primarily in the 171~{\AA} passband (Figures~\ref{1104_aia}(d)--(e) and \ref{1104_ha}(c), (e)). 

When the upper threads in the right section of the filament rise, new thread-like structures are illuminated in the AIA 131~{\AA} passband (from $\sim$13:22~UT), which connect the cool threads with several patches of positive photospheric flux near the left end of the filament (Figure~\ref{1104_aia}(c)--(d)). The photospheric polarities in the end regions of the rising structures imply a sinistral chirality for the filament in agreement with the handedness of the indicated twist. The rising hot threads form a hot channel, which soon extends along the whole filament and whose subsequent fast rise and eastward extension coincide with the impulsive flare phase (Figure~\ref{1104_aia}(e)--(g)). This and the exclusive visibility in the AIA passbands sensitive to plasma at temperatures above 6~MK are typical for a hot channel \citep{cheng11_fluxrope,cheng13_driver,cheng16_apjs,zhang12,patsourakos13,tripathi13,aparna-tripathi16}. 

Simultaneously with the rise of the hot channel, the flare ribbons and flare loop arcade begin to extend along the whole filament and even beyond its eastern end. Both become brightest along the left section of the filament (Figures~\ref{1104_aia}(f)--(g) and \ref{1104_ha}(e)). This also becomes the location of the main HXR sources, which start at scattered locations between the hot channel and remaining filament threads but intensify strongest near the left section (Figure~\ref{1104_aia}(e)--(f)). Eventually, an HXR loop-top and a pair of HXR footpoint sources are formed here (Figure~\ref{1104_aia}(g)). The filament remains visible in this area (Figures~\ref{1104_aia}(e)--(g) and \ref{1104_ha}(d), (f)), implying a vertical split and partial eruption along this section of the filament channel as well. The flare ribbons, flare loop arcade, and HXR sources demonstrate that the reconnection associated with the vertical split develops seamlessly into the standard flare reconnection in a vertical current sheet in the impulsive flare phase. 

Different from the right section of the erupting flux, the splitting in the left section occurs above the filament. This is not inconsistent with the assumption that a single MFR has formed along the whole filament channel. The appearance of the filament threads in the right section suggests that they are located close to the magnetic axis of the MFR, winding below as well as above the axis. The threads in the left section might all run below the axis, consistent with their large-scale forward-S shape. 

The vertical split of the coronal magnetic flux in the filament channel is well developed by $\sim$13:30~UT (Figures~\ref{1104_aia}(d) and \ref{1104_ha}(c)), i.e., before the onset of the impulsive flare phase. The clear development of the split, as well as the simultaneous formation of the hot channel from $\sim$13:22~UT, occur during a relatively impulsive SXR enhancement of $\sim$10~min duration at the $\sim$C1.5 level, which is more of the nature of a flare precursor than a slow-rise phase. The X-ray light curves in Figure~\ref{1104_aia}(a) imply a seamless transition from this precursor to the impulsive flare phase, similar to the spatial characteristics of the reconnection signatures discussed above. The interval $\sim$13:15--13:22~UT may be considered as a slow-rise phase of the filament.

Furthermore, the reconnection associated with the vertical splitting indicates a relationship with the flux cancellation in the filament channel. Figure~\ref{1104_mag} shows the temporal evolution of the spatially integrated 1600~{\AA} intensity and the positive and negative magnetic fluxes at the cancellation site under the first brightenings. Here the total magnetic flux continuously decreases from $\sim$13:00 UT, with a drop of 80\% (30\%) in the negative (positive) polarity by 13:40~UT. In the same period, the intensity of the EUV/UV brightenings shows three enhancements, the stronger ones commencing at 13:15~UT and all reaching their peak before the impulsive flare phase. Thus, the temporal characteristics show a continuous evolution from the slow tether-cutting reconnection associated with flux cancellation to the faster reconnection associated with the splitting of the flux, which, in turn, evolves seamlessly into the fast flare reconnection of the main event. 

However, the spatial characteristics of the cancellation-associated reconnection and the splitting-associated reconnection are likely different. The former is assumed to operate in the very low atmosphere, predominantly at photospheric and chromospheric levels \citep{vanballegooijen89}, whereas the latter operates in the low corona. Correspondingly, the latter has a relatively impulsive onset ($\sim$5~min rise time) in the precursor phase of the flare, which is much shorter than the cancellation time scale of 20--30~min in Figure~\ref{1104_mag}(c), even though only a small cancellation area is considered there. The difference in time scale indicates that the split is ultimately caused by the onset of a fast coronal process, i.e., a loss of equilibrium (instability) of the rising flux. This interpretation is also supported by the fact that a hot channel has been built up in the precursor phase. A hot channel is generally seen in the corona during the periods of enhanced reconnection, typically in the precursor phase of major eruptions. Slow flux cancellation in a short period is not efficient to produce a hot channel, which usually needs a longer time \citep[e.g., hours to days;][]{tripathi09,cheng14_formation}.

\begin{figure}
\center {\includegraphics[width=8cm]{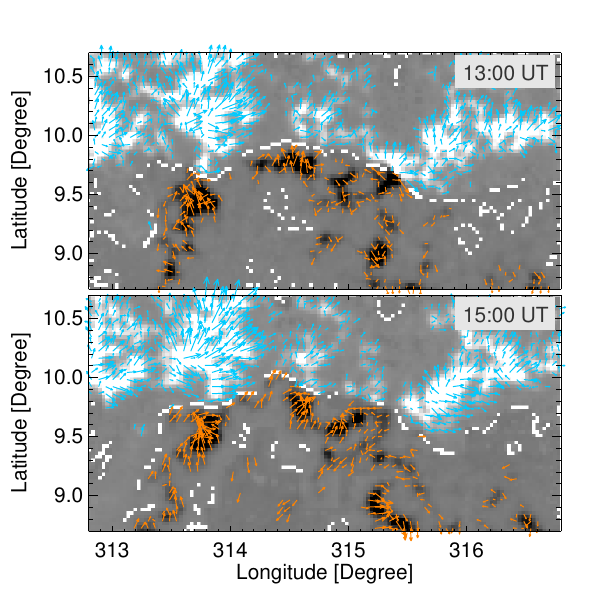}}
\caption{HMI vector field before and after the partial filament eruption. The background image and arrows show the vertical and transverse components of vector field, respectively. The white segments indicate the locations of BPs.}
\label{1104_bp}
\end{figure}

We also inspect the photospheric vector data in the region of interest before and after the eruption provided by the HMI Active Region Patches (HARPs), in which the transverse field 180 degree ambiguity has been corrected via the minimum energy method \citep[][Figure~\ref{1104_bp}]{metcalf06,bobra14}. In the magnetogram taken at 13:00~UT, many magnetic field vectors at the PIL point in the inverse direction, from the negative to the positive polarity, showing the appearance of BPs. We compute the locations of BPs (the white segments in Figure~\ref{1104_bp}) using the definition, $(\mathbf{B} \cdot \nabla) B_z > 0$ and $B_z=0$, as given in \citet{titov93}. One can clearly see that the BPs appear at almost the whole PIL before the eruption, demonstrating that the bottom of the pre-eruptive filament-hosting MFR touches the photospheric surface in many points. Moreover, it is seen that a majority of BPs remain after the eruption. This agrees with the EUV and H$\alpha$ observation that the filament splits vertically and its lower part remains in the original place. The magnetic field line tying in the BPs prevents the lower part of the MFR from erupting and only allows its upper part (which becomes the hot channel) to erupt through vertical splitting and internal reconnection, as conjectured in \citet{gibson06_apjl}. 

\subsubsection{Kinematics of the Eruption} 

\begin{figure*}
\center {\hspace{-0.07\textwidth} \includegraphics[width=15cm]{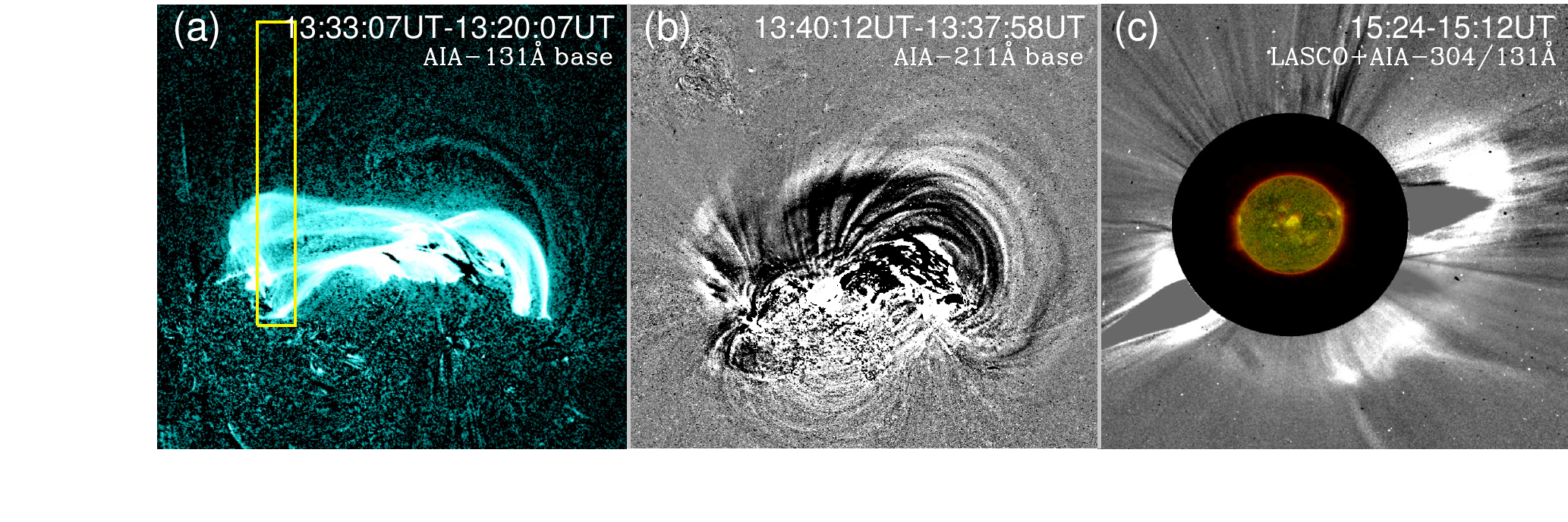}\vspace{-0.03\textwidth}}
\center {\includegraphics[width=16.2cm]{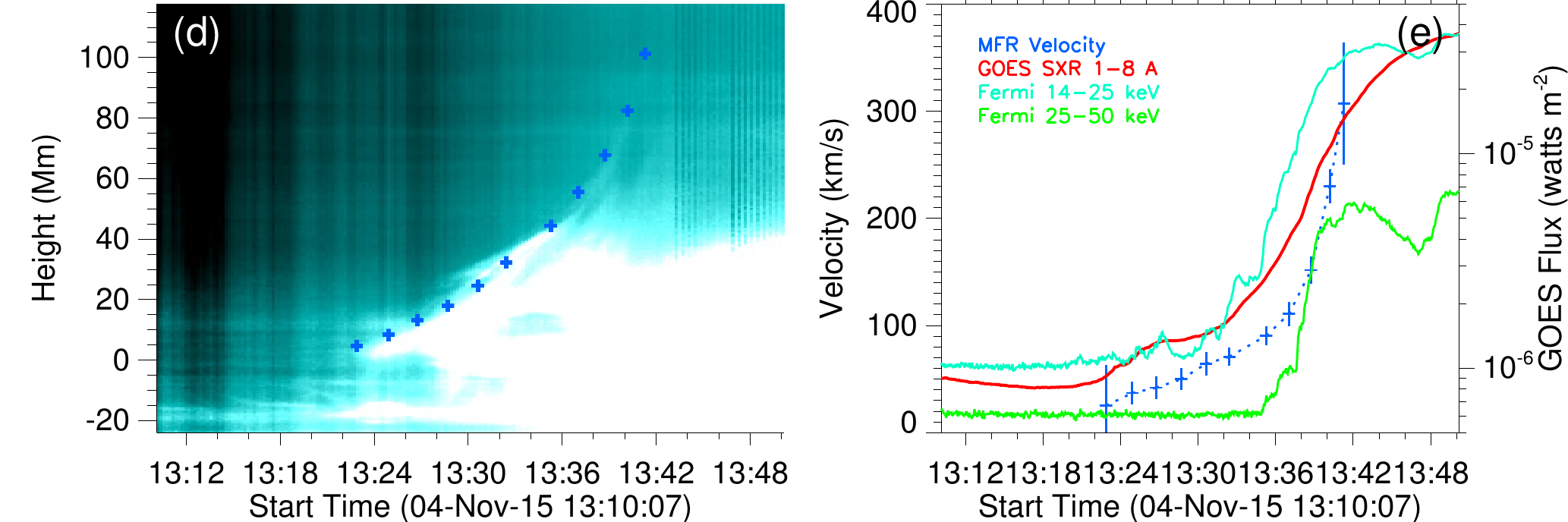}}
\caption{(a) and (b) AIA 131 {\AA} and 211 {\AA} base-difference images showing the erupting hot threads and the CME. The vertical rectangle in Panel (a) marks the selected slice along the direction of the eruption. (c) LASCO/C2 white-light base-difference image overlaid by the composite image of the AIA 131 {\AA} and 304 {\AA} passbands. (d) Slice-time plot of the AIA 131 {\AA} images with the blue pluses showing the height-time measurements of the hot channel in the partial  eruption. (e) Temporal evolution of the eruption velocity (blue), \textsl{GOES} SXR 1--8 {\AA} flux (red), and \textsl{Fermi} 14--25 and 25--50 keV fluxes (cyan and green, respectively).}
\label{1104_ht}
\end{figure*}

Different from the 2014 December 24 event, the 2015 November 4 event shows a precursor in the SXR flux (from $\sim$13:22~UT) which may form a hot channel and the escape of the hot channel as a CME (Figure~\ref{1104_ht}). To investigate the kinematics of the ejected flux, we put a vertical slice along the direction of the eruption in the AIA 131~{\AA} images (yellow rectangle in Figure~\ref{1104_ht}(a)). Figure~\ref{1104_ht}(d) shows the slice-time plot and the height-time measurements (blue pluses). One can see that the hot channel rises immediately from its formation at $\sim$13:22~UT. (Note that the dark threads in the right section of the filament rise slowly already from $\sim$13:15~UT.) The rise is accompanied by the enhancement of the EUV brightenings underneath the hot channel. After $\sim$13:40~UT, due to the quick expansion, the hot channel gradually fades.

The kinematic evolution of the hot channel is further shown in Figure~\ref{1104_ht}(e). It is seen that the velocity increases slowly and approximately linearly from near zero to $\sim$70~km~s$^{-1}$ during $\sim$13:22--32~UT, corresponding to an acceleration of $\sim$120~m~s$^{-2}$. This phase of moderate acceleration corresponds in time to the precursor enhancement of the flare X-ray emission as seen in the \textit{GOES} 1--8~{\AA} and \textit{RHESSI} 6--12~keV light curves. 

The simultaneous onset of reconnection signatures in the present event, as EUV brightenings in the slow-rise phase and as enhanced SXR flux in the precursor, differs from the delayed onset of reconnection signatures in the confined event of Section~\ref{ss:confined}. This difference is presumably due to the different height of the split relative to the main body of the filament. The splitting in the confined event proceeds under most of the filament, whereas it cuts through the main body of the filament's right section here. Therefore, the splitting leads to immediate brightenings in this event, while in the confined event, the brightenings become significant only when the vertical current sheet of the subsequent flare is established.

After $\sim$13:32~UT, the hot channel is impulsively accelerated, showing an amplifying acceleration. This is closely synchronized with the fast, main rise of the \textit{Fermi} 14--25 and 25--50~keV fluxes, which indicate the onset of fast flare reconnection. The velocity increases from $\sim$70~km~s$^{-1}$ at $\sim$13:32~UT to $\sim$300~km~s$^{-1}$ at $\sim$13:42~UT with an average acceleration of $\sim$400~m~s$^{-2}$. The synchronization in this impulsive flare phase reflects the usual mutual feedback between the rise of unstable flux and fast flare reconnection in a vertical current sheet underneath. 

\section{Summary and Discussion}\label{s:Summary}

In this paper, we study two partial eruptions, the first of which appears as a confined filament eruption, while the second is a successful eruption (evolves into a CME). Thanks to the high resolution, high cadence, and multi-temperature capability of the AIA, both partial eruptions are found to be a result of the vertical splitting of a filament-hosting MFR involving a loss of equilibrium and internal reconnection. 

At the onset of the first event (following a short slow-rise phase), a large part of the threads begin to rise along most of the filament. The first EUV brightenings, located between the rising and remaining threads, and the onset of the associated SXR flare occur only $\sim$4~min later. The resulting pattern of twisted dark and bright threads indicates that an MFR exists already at the onset of the activity. Since the fast partial rise of the filament commences suddenly and significantly precedes the signatures of reconnection, this event clearly favors an ideal MHD loss of equilibrium as the onset and drive mechanism. The strong writhing and the confinement of the erupting filament suggest the helical kink mode, but the data do not allow inferring this mode unambiguously, so that a torus instability is also possible. The formation of the HXR sources at the positions of the strongest EUV brighenings indicates that the internal reconnection involved in the splitting seamlessly evolves into the fast flare reconnection in a vertical current sheet. These observations are perfectly consistent with the model for partial eruptions by \citet{gibson06_apjl}, which assumes the BPS topology for an unstable MFR. The line tying in the BP prevents the lower part of the MFR from taking part in the eruption, so that an instability of the upper part enforces a vertical split which is completed by internal reconnection. Unfortunately, a reliable search for BPs at the position of the filament is not possible, due to its considerable distance from the central meridian.

The second event begins with a slow rise of the filament's right section and simultaneous EUV brightenings in the volume of the rising section. Subsequently, the indications of internal reconnection associated with the vertical splitting of this part of the filament are very similar to the first event: amplified EUV/UV brightenings and the formation of coronal HXR sources between the erupting and remaining flux. A vertical split of magnetic flux in the filament channel occurs here along the whole filament, but cuts through the filament only in its right section, whereas the left section must be completely situated in the lower, remaining flux and is not significantly affected. The threads in the right section of the filament indicate twist from the onset of the brightenings, i.e., a pre-existing MFR. In this event, the first signs of filament splitting and the first brightenings (due to reconnection) appear simultaneously in a slow-rise phase of $\sim$7~min duration and amplify simultaneously in a subsequent precursor phase of $\sim$10~min duration. These are followed seamlessly by the onset of the main upward acceleration and impulsive flare phase, which are also synchronized and lead to standard flare ribbons and loops. The event is preceded by a long period of flux cancellation under the filament; in particular, a strong cancellation episode is closely associated with the first brightenings in space and time. While the cancellation is the prime candidate for the driving of the region toward the onset of the eruption, the impulsive onset of the precursor and the prompt formation of a hot channel indicate that a faster coronal process sets in and drives the splitting---a loss of equilibrium in the erupting flux. The vector magnetogram shows the inverse magnetic field direction, corresponding to BPs, along most of the PIL under the partially erupting filament. Most BP sections are also present after the eruption. Thus, this partial eruption exhibits \emph{all} features of the model by \citet{gibson06_apjl}. 

Both events do not support an interpretation based on the double-decker MFR configuration. For the most likely case that both MFRs in this configuration possess the same direction of the axial field and the same handedness, there is an X-type magnetic structure (a hyperbolic flux tube) between the two MFRs. This is the natural place for the split to happen, and it would readily collapse into a vertical current sheet \citep{titov03}. Therefore, neither the internal split of the filament, as observed in the successful eruption, nor the delayed onset of reconnection signatures, as observed in the confined eruption, are expected for the double-decker MFR configuration. Actually, the source region of the successful eruption showed an indication for a second MFR lying above the filament during the interval $\sim$8.5--2.5~hr prior to the onset of the eruption. This was indicated by the transient appearance of a second filament which faded and was then replaced by a transient hot loop \cite[see images in][]{wang17_nc}. It is unclear whether a second MFR still exists above the filament by the time of the partial eruption. If so, it is not directly involved in the vertical split which occurs \emph{within} the right part of the filament.

The partial eruptions also shed a new light on homologous eruptions. It has previously been suggested that homologous eruptions may originate from successive eruption of a long magnetic flux system or different magnetic flux systems \citep{zhangjun02,wangym13}. However, the partial eruptions induced by the MFR splitting suggest that homologous eruptions could also result from multiple partial eruptions of a single MFR. After part of the flux is released, there remains a considerable free energy in the surviving flux \citep{gibson06_apjl}, which can probably initiate a second eruption in a short time \citep[e.g.,][]{yangkai16}. Recently, the multiple eruption of an MFR has been demonstrated numerically by \citet{chatterjee13}, however, in this study, the surviving MFR was continually energized through magnetic flux emergence. 

Finally, it should be mentioned that besides the internal reconnection in the wake of the erupting flux, magnetic reconnection can also take place between the erupting and ambient flux, for example, due to a change in the direction of the erupting flux being writhed. In the confined filament eruption studied here, the writhing and external reconnection are well observed. External reconnection of a writhing erupting filament has also been found in numerical simulations \citep{shiota10,hassanin16,hassanin16a}.

\acknowledgements We thank the referee Durgesh Tripathi for useful suggestions. We also thank Linhua Deng for providing us the NVST data, Rui Liu for helpful discussions, and the ISSI workshop on ``Decoding the Pre­-Eruptive Magnetic Configuration of Coronal Mass Ejections" led by S. Patsourakos \& A. Vourlidas. \textit{SDO} is a mission of NASAs Living With a Star Program. X.C. and M.D.D. are supported by NSFC under grants 11722325, 11733003, 11790303 and by Jiangsu NSF under grants BK20170011. X.C. is also supported by ``Dengfeng B" program of Nanjing University. B.K. acknowledges support by NASA under Grants NNX16AH87G and 80NSSC17K0016, the NSFC, and the DFG.

\end{document}